%% file: tsamis_m42.tex
\documentclass[vecphys]{svmult}

\usepackage{makeidx}         
\usepackage{graphicx}        
\usepackage{multicol}        

\makeindex             

\begin{document}

\title*{Integral field spectroscopy of protoplanetary disks in Orion with VLT FLAMES}
\titlerunning{VLT FLAMES mapping of proplyds in M42}
\author{Y. G. Tsamis\inst{1}, J. R. Walsh\inst{2} \and
D. P\'{e}quignot\inst{3}}
\institute{Department of Physics and Astronomy, University College London,
      Gower Street, London WC1E 6BT, U.K.
\texttt{ygt@star.ucl.ac.uk}
\and Space Telescope European Co-ordinating Facility, ESO, Karl-Schwarzschild Strasse 2, D-85748 Garching, Germany
\texttt{jwalsh@eso.org} \and LUTH, Observatoire de Paris, CNRS, 5 Place
J. Janssen, 92190 Meudon, France \texttt{Daniel.Pequignot@obspm.fr}}
%
%
\maketitle

\begin{abstract}
We discuss integral field spectroscopy of proplyds in M42 using the FLAMES
Argus unit and report the first detection of recombination lines of C~{\sc ii}
and O~{\sc ii} from the archetypical LV2 object. These lines can provide
important new diagnostics of the physical conditions in proplyds. We also draw
attention to the future capabilities of the MUSE spectrograph in relation to
similar studies.
\end{abstract}

\section{Rationale}
\label{sec:1}

This contribution focuses on optical integral field spectrophotometry (IFS) of
protoplanetary disks (proplyds) in M42 taken with the Argus unit of VLT FLAMES.
The proplyds in M42 are partially ionized, low-mass embedded young stellar
objects immersed in the extreme UV radiation field of the Trapezium cluster
\cite{Bally}. They represent a unique nearby environment for the study of
ongoing star formation in a region dominated by main sequence OB-type stars. At
optical wavelengths proplyds usually present a photoionized skin facing the
Trapezium, giving way to a neutral dusty envelope which is shielded from the
ionizing photons and is often shaped into `cometary' radiation-bounded tails.
IFS mapping of the proplyds and their immediate surroundings can yield new
insight on the influence of small-scale `inhomogeneities' on integrated spectra
of distant galactic and extragalactic H~{\sc ii} regions.

With this programme we specifically aimed at recording the faint optical
recombination line (ORL) spectra of carbon (C~{\sc ii}) and oxygen (O~{\sc ii})
ions and use them, for the first time, as abundance diagnostics of the
photoionized surfaces of the proplyds and their outflows. Simultaneous coverage
of the strong collisionally excited lines (CELs) of [O~{\sc iii}] would allow
us to obtain a separate estimate of the oxygen abundances across the field and
thus study the `abundance discrepancy problem', whereby heavy element
abundances, relative to H, from ORLs are found to be higher than the
corresponding CEL-based abundances for classic H~{\sc ii} regions such as M42
and 30 Doradus, by factors of up to $\sim$ 2 \cite{T03}. The resolution of this
problem is of high priority as it casts uncertainty on classic CEL-based
methods of abundance determinations for local and distant nebulae and galaxies.
It has been proposed that temperature fluctuations \cite{P03}, density
fluctuations, or zones of hydrogen-deficient plasma \cite{TP05} within the
nebulae may contribute to the `ORL vs. CEL' problem. Our observations were
aimed at discriminating between the various possibilities.


\section{Observations and data reduction method} \label{sec:2}

Argus is a rectangular array of 22$\times$14 microlenses fed by optical fibres:
we used the small configuration which provides a sampling of 0.30
arcsec$^2$/microlens and projects 6.6$\times$4.2 arcsec$^2$ on the sky,
yielding 297 positional spectra per field of view. The targets were three
relatively bright proplyds, including the archetypical object LV2 (167-317;
\cite{LV}), and were selected from {\it HST} WFPC2 H$\alpha$ and [O~{\sc iii}]
images of the nebula. The spectra were taken in service mode, and in sub-arcsec
seeing, with the LR1--5 and HR1, 3, 4, 6, 8, 14B grating set-ups of the Giraffe
spectrograph at resolving powers of $\sim $ 10\,000 -- 46\,000. For LV2 the
total exposure time in the LR/HR modes was 3000 and 5900 sec respectively;
similar times were allocated to the other two targets. The data reduction was
done with the girBLDRS pipeline developed by the Geneva Observatory.
Custom-made {\sc iraf} scripts allowed us to construct data cubes and spectral
line maps, and a dedicated $\chi^2$ minimization routine was used to
automatically fit Gaussians to the emission lines.

\section{First results}

At the time of writing preliminary monochromatic maps have been obtained for
LV2. We succeeded in detecting the C~{\sc ii} $\lambda$4267 and O~{\sc ii}
$\lambda$4649 ORLs from the head of the proplyd, and imaged the proplyd head
and outflow in the light of H$\alpha$, and the [O~{\sc iii}]
$\lambda\lambda$4363, 4959, [Ar~{\sc iv}] $\lambda\lambda$4711, 4740, and
[S~{\sc ii}] $\lambda\lambda$6716, 6731 CELs. The [O~{\sc iii}] line ratio
yielded an electron temperature ($T_{\rm e}$) map and the [Ar~{\sc iv}] and
[S~{\sc ii}] ratios yielded electron density ($N_{\rm e}$) maps. In Fig.\,~1
spectra of LV2 are shown extracted over the proplyd's tip ($\sim$ 9 spaxels).
The C~{\sc ii} $\lambda$4267 3d--4f line is well detected in the LR2 setting,
as are the numerous O~{\sc ii} V1 multiplet 3s--3p lines around 465.0\,nm in
the HR6 setting. All these lines are useful abundance diagnostics \cite{TW07},
and the intramultiplet relative intensities of O~{\sc ii} V1 lines are a
$N_{\rm e}$ diagnostic of their emitting regions too \cite{BS06}.

\begin{figure}
\centering
\includegraphics[scale=0.46, angle=-90]{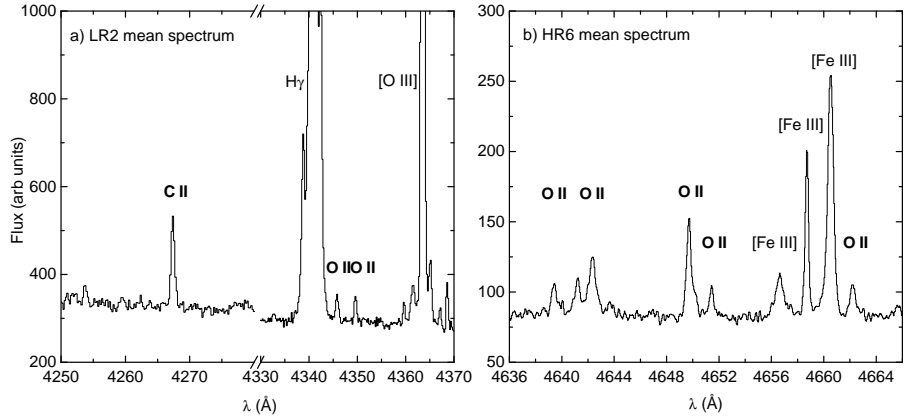}
\caption{LV2 (167-317) Argus spectra: a) LR2 spectrum showing the C~{\sc ii}
$\lambda$4267 recombination line; b) HR6 spectrum showing the O~{\sc ii} V1
multiplet recombination lines. Note the velocity splitting of the [Fe~{\sc
iii}] $\lambda$4658 line with the red- and blue-shifted components arising from
the proplyd's outflow. }
\label{fig:1}       
\end{figure}

In Fig.\,~1b the O~{\sc ii} lines appear single peaked meaning that they mostly
arise at the LV2 rest frame velocity: in contrast, the [Fe~{\sc iii}]
$\lambda$4658 line exhibits two additional velocity components associated with
the proplyd's bipolar outflow; further detections of [Fe~{\sc iii}] lines,
which are good tracers of shocked gas, will allow us to measure the electron
density in the outflow. Analysis of the [O~{\sc iii}] $\lambda$5007 line
substructure at a spectral resolution of 9\,km s$^{-1}$ pix$^{-1}$ indicates
that the lobes of the bipolar jet emanating from the embedded protostar have
line of sight velocities of approximately $-$100 and $+$80\,km s$^{-1}$ (see
Fig.\,~2). It is likely, however, that the velocity structure will differ
amongst various ionic species; a detailed kinematical analysis of emission
lines from the numerous Giraffe HR settings will clarify this. The velocity
resolution of this dataset at H$\alpha$ is a factor of 2.3 higher compared to a
Gemini-S GMOS IFU analysis \cite{V05} (which focused only on the red part of
the optical spectrum and did not go deep enough to detect any heavy element
ORLs).

\begin{figure}
\centering
\includegraphics[scale=0.4, angle=-90]{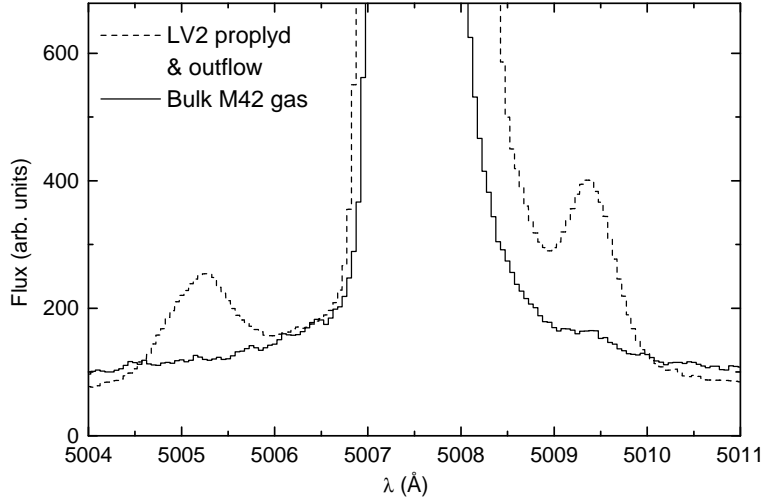}
%
\caption{LV2 (167-317) Argus spectrum showing the [O~{\sc iii}] $\lambda$5007
line taken with the HR8 setting (at 9\,km\,s$^{-1}$\,pix$^{-1}$); note the
blue- and red-shifted components arising from the proplyd's outflow and which
are absent in the background nebula.}
\label{fig:1}       
\end{figure}

In Fig.\,~3 we show a $T_{\rm e}$([O~{\sc iii}]) map of LV2 and its immediate
vicinity based on the $\lambda$4363/$\lambda$4959 ratio with the corresponding
O$^{2+}$/H$^+$ CEL abundance map based on the $\lambda$4959 line. Notably, the
electron temperature over the proplyd appears to be as high as 15\,000\,K
whereas the background temperature is only $\sim$ 8500\,K. This is mainly due
to collisional suppression of the relatively low critical density $\lambda$4959
line over the dense proplyd (at $N_{\rm e}$ $>$ 7$\times$10$^5$\,cm$^{-3}$). As
a direct result of such pseudo-$T_{\rm e}$ fluctuations, the O$^{2+}$/H$^+$ CEL
abundance (Fig.\,~3 right) in the close vicinity of LV2 appears to be about a
factor of 3 lower than in the background nebula.

An intricate combination of pseudo-variations in CEL-based abundances
\emph{coupled with} enhanced `metallic' ORL emission from dense proplyds, and
similar types of condensations, could at last provide a factual explanation to
the long-standing abundance discrepancy problem in H~{\sc ii} regions. Detailed
studies with VLT FLAMES can improve our understanding of the protoplanetary
disks themselves and help elucidate the influence of such objects on integrated
nebular spectra.

In conclusion, we note that the planned second generation VLT instrumentation
will include MUSE, an adaptive optics assisted IFU spectrograph, which in wide
field mode (1$\times$1 arcmin$^2$) will be able to take integral field spectra
of the whole central Orion region (and of distant giant H~{\sc ii} regions),
and to sample simultaneously the entire proplyd population with
0.2$''$$\times$0.2$''$ spatial pixels. In narrow field AO-assisted mode the
spatial resolution of 0.025$''$$\times$0.025$''$ will allow unprecedented views
of individual proplyds and protostellar outflows. It is unfortunate, however,
that the currently planned wavelength coverage of MUSE is only 465.0 -- 930.0
nm; this will \emph{miss} the brightest optical recombination lines from C~{\sc
ii}, C~{\sc iii}, N~{\sc ii}, N~{\sc iii}, O~{\sc ii}, Ne~{\sc ii} species ($<$
465.0 nm) which have in recent years opened an exciting new window to the
chemistry and astrophysics of nebulae.

\begin{figure}
\centering

\includegraphics[scale=0.40, angle=0]{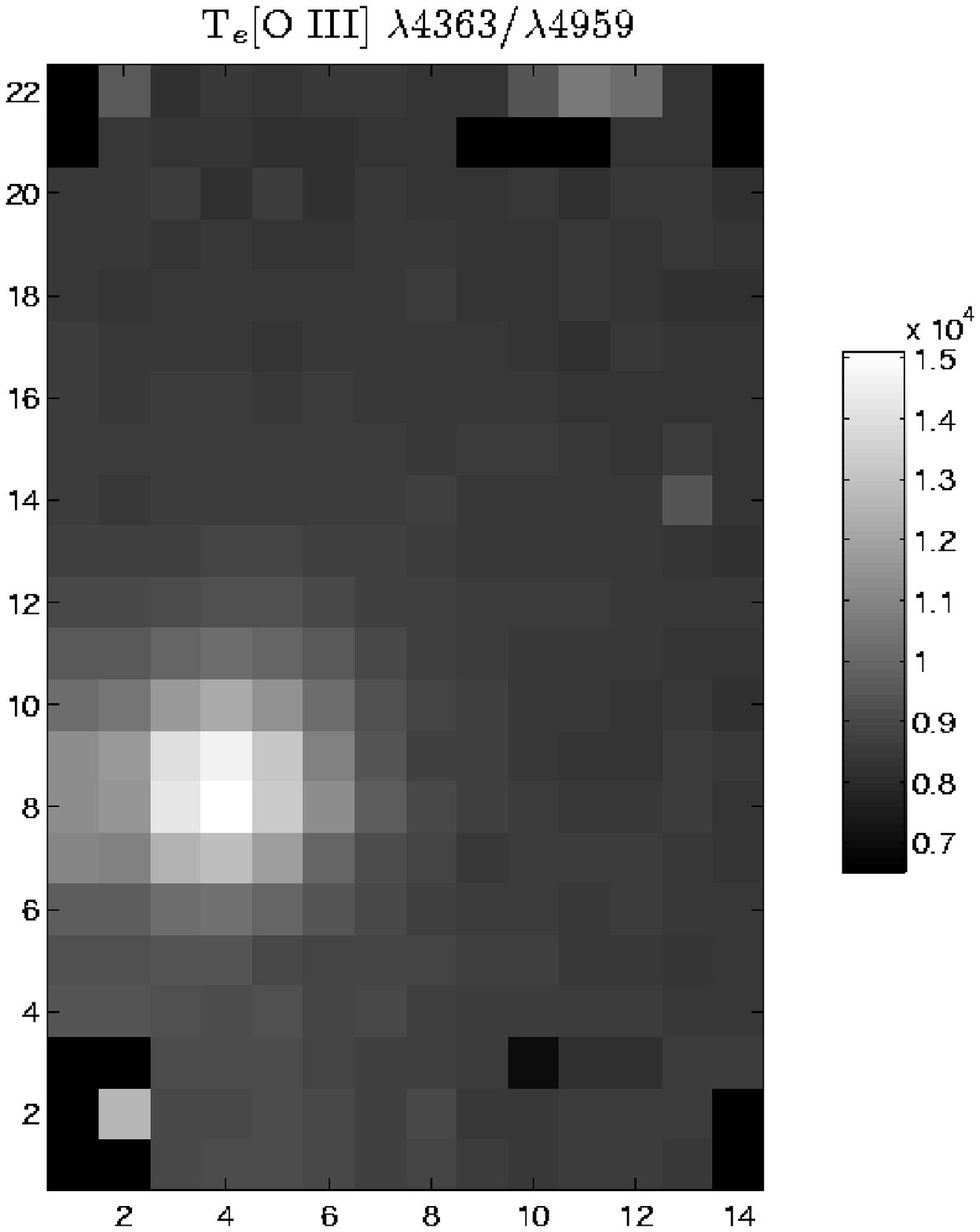}
\includegraphics[scale=0.40, angle=0]{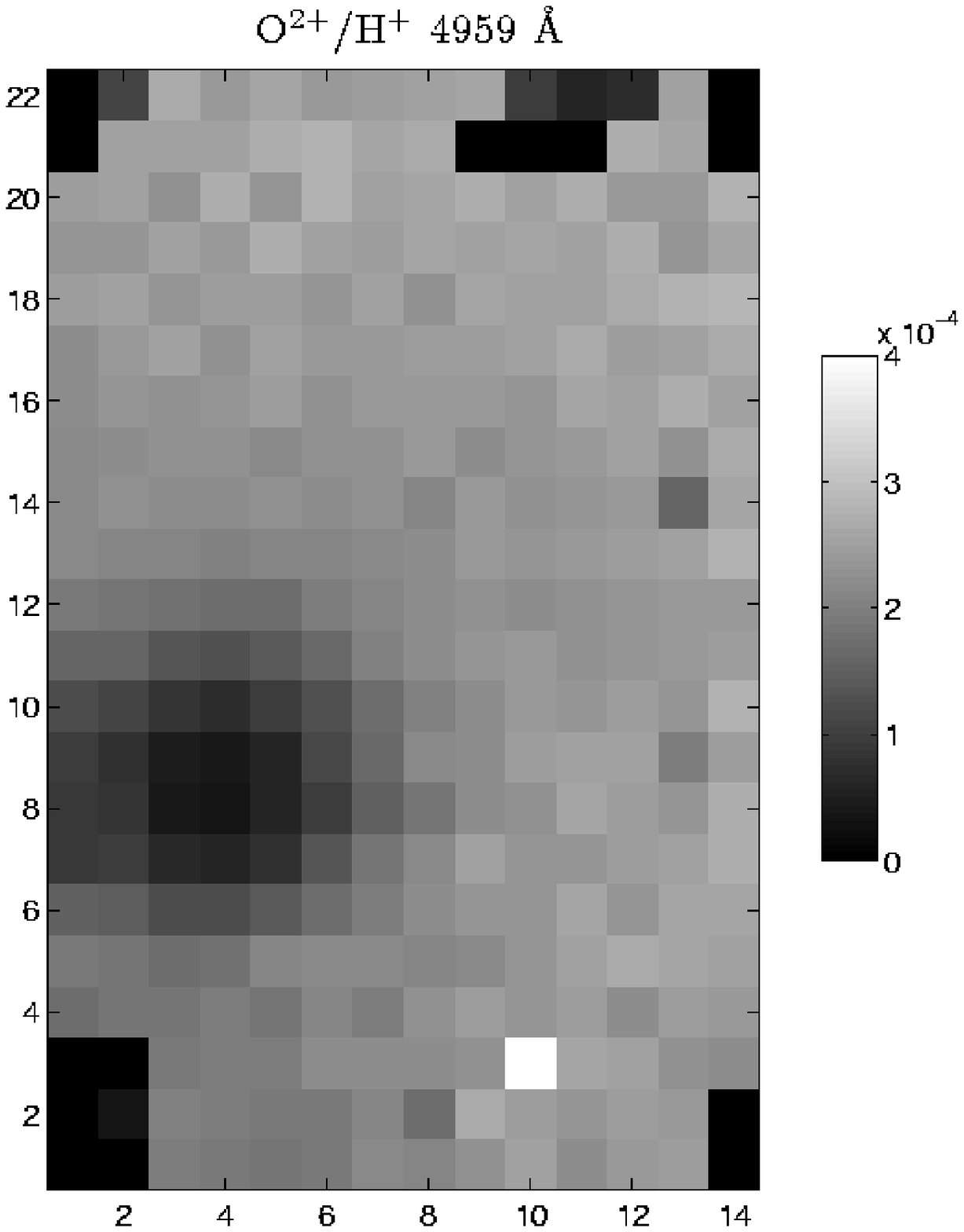}


\caption{LV2 (167-317) and M42 background physical conditions: (Left) The
electron temperature map based on the [O~{\sc iii}] auroral to nebular line
ratio; (Right) The corresponding forbidden-line doubly ionized oxygen abundance
map.}
\label{fig:1}       
\end{figure}

%
\input{referenc}



\end{document}

%% file: referenc.tex
%
%

%
%